# Sedimentation of particulate suspensions under stagnant conditions in horizontal pipes

Tanmoy Das[a], Daniel Lester[a], Anthony Stickland[b], Nicky Eshtiaghi[a]

[a] School of Engineering, RMIT University, Melbourne, Victoria 3001, Australia

[b] ARC Centre of Excellence for Enabling Eco-Efficient Beneficiation of Minerals, Department of Chemical Engineering, The University of Melbourne, Victoria 3010, Australia

**Abstract**

Sedimentation of particulate suspensions in horizontal pipes can lead to formation, growth and consolidation of a solid-like bed which can severely retard pipeline performance. As stagnant flow conditions frequently arise during industrial processes, critical operational questions are: (i) at what rate and extent does sedimentation proceed, and (ii) can the sedimentation dynamics be predicted from conventional suspension characterisation methods? We address these questions by characterising the sedimentation properties of an aqueous Kaolin suspension via batch settling tests and comparing predictions from 1D sedimentation theory with experiments in a horizontally oriented cylindrical pipe. We show that particulate sedimentation can be accurately predicted, indicating that the estimated sedimentation properties are representative material properties, and that transient effects such as gravity currents are not significant. Conversely, we find that the consolidation of the sediment is not well predicted by 1D theory, suggesting that the stress state is not 1D and likely involves contributions from the pipe walls. These stagnant cylindrical pipe results provide a basis for the development of methods to predict pipeline sedimentation under more general (laminar and turbulent) flow conditions.

**Keywords:** sedimentation, pipeline transport, settling velocity

**1. Introduction**

Transportation of particulate suspensions via pipeline systems is a common process in the water, chemical, mineral, food and petroleum industries (Peker and Helvaci, 2008). Although such suspensions typically remain well-mixed under turbulent flow, under stagnant or laminar flow conditions, the constituent phases of these suspensions can separate over time due to their density difference, leading to the formation and consolidation of a solid-like bed at the bottom of a horizontal pipe. With time, accumulation of solids in this bed can reduce pipeline capacity and efficiency, and lead to pumping instabilities and ultimately pipe blockage. Procedures for removing solids beds or unblocking the pipes leads to production losses, labour and materials costs and the potential for contamination of the environment (Kotzé et al., 2019). Thus, understanding the settling behaviour of particulate suspensions in horizontal cylindrical pipes is critically important in order to develop pipeline design tools and operating procedures to effectively manage solids deposition and potential pipe blockages (Cunliffe et al., 2021).

Under sufficiently turbulent pipe flow conditions, dense particles may be suspended via turbulent dispersion forces, and at very high flow rates these particles are near-homogeneously distributed with a near-uniform average solids volume fraction distribution (Peker and Helvaci, 2008). At intermediate flow rates, the solid particles may remain suspended, but the solids concentration distribution is non-uniform, reflecting competition between gravitational settling and turbulent dispersion. With further decrease in the average suspension velocity, fully suspended flow changes to two-layer flow (Wilson, 1970). At even lower flow rates, a stationary bed will form at the bottom of the pipe. The transition between these flow patterns is usually determined by visual observations and defined by the limit deposit velocity (Durand, 1953; Newitt and Richardson, 1955) which is the minimum velocity required to maintain a fully suspended flow.

When flow ceases in a pipe containing a suspension, the particles will sediment over time if the gravitational forces are sufficient to overcome Brownian motion, leading to build-up of a consolidated bed of particles. If these particles are cohesive, this bed will exhibit a finite shear yield stress that must be overcome to for the bed to start to deform, a necessary precursor to fluidization and resuspension. The longer the pipe is left under stagnant conditions, the greater the extent of consolidation (even if sedimentation has completed), further increasing the shear yield stress profile throughout the bed. Hence the resistance of cohesive beds to resuspension depends intimately upon the sedimentation and consolidation processes.

Although much attention has been paid to the bed fluidization and resuspension process, much less is known regarding the sedimentation and consolidation process under stagnant flow conditions. In this study we address this problem by considering the dynamics of sedimentation and consolidation in horizontally oriented pipes under stagnant flow conditions. We analyse these flows via the phenomenological 1D theory (Buscall and White, 1987) developed for the sedimentation and consolidation of colloidal suspensions, which has been adapted for vessels with varying cross-sectional area as a function of height (Anestis, 1981; Landman and White, 1994; Bürger et al., 2018; Bürger et al., 2004) such as horizontally-oriented cylinders. Of primary importance is the question whether suspension properties measured via vertical cylinder tests can be used to accurately predict sedimentation and consolidation in horizontal pipes under stagnant flow conditions.

In this study we address this shortcoming by performing a series of sedimentation tests in vertical and horizontal cylinders under stagnant flow conditions. Gravity batch settling tests in vertical cylinders are advantageous to those conducted in horizontal cylindrical pipes as the former requires a simple experimental setup, and analysis of data is simpler than for the curved cross-sectional geometry that arises in the horizontal case. We measure the sedimentation rate

(via evolution of the suspension/supernatant interface) of three aqueous Kaolin suspensions at the isoelectric point with different initial solids concentrations (1, 1.5 and 2.5 vol%) in settling tests in both vertically- and horizontally-oriented cylinders. Sedimentation data from the vertical cylinder test with 1 vol% initial solids concentration was used to estimate the material properties that govern sedimentation. The material properties are then used with numerical simulation to predict sedimentation in horizontally oriented cylinders, the results of which were compared with the experimental measurements. Successful validation of these numerical predictions would verify that conventional batch settling tests are a valid method to predict sedimentation of particulate suspensions in horizontal pipes under stagnant flow conditions. This study also acts an important stepping stone toward prediction of particulate sedimentation under laminar and turbulent pipe flow,

The remainder of this paper is structured as follows. In Section 2 we briefly outline the theory governing gravity separation of non-Brownian particulate suspensions in vessels of arbitrary cross-sectional profile, materials and methods are outlined in Section 3. Results are presented and discussed in Section 4, followed by Conclusions in Section 5.

## 2. Theory
### 2.1 Separation of non-Brownian particulate suspensions

The basic separation behaviour of non-Brownian particulate suspensions has been successfully described by many workers (Buscall and White, 1987; Fitch, 1979; Kynch, 1952; Michaels and Bolger, 1962) via a phenomenological 1D theory of sedimentation and consolidation that quantifies both the sedimentation of dense colloidal particles under reduced hydrodynamic mobility and the consolidation of a particulate network under an applied compressive stress. At solids concentrations (quantified by the local average solids volume fraction $\phi$) greater than a critical value termed the gel point $\phi_g$, the solids phase forms a poro-elastic network (particulate gel) that can withstand and transmit compressive stress, signifying the transition from sedimentation to consolidation behaviour. Under this 1D theory, the basic separation properties of particulate suspensions are characterized in terms of the compressive yield stress $P_y(\phi)$ (with $P_y(\phi) = 0$ for $\phi < \phi_g$) which characterises the compressive strength of the colloidal gel, and the hindered settling function, $R(\phi)$ which characterises the interphase hydrodynamic drag from dilute to highly concentrated regimes. Both of these material functions strongly increase with solids concentration. They provide the ability to predict separation of particulate suspensions in a variety of 1D applications such as batch settling, gravity thickening, filtration and centrifugation (Stickland, 2015; Stickland and Buscall, 2009).

For many industrial applications, the validity of the assumption of 1D separation behaviour is an open question, and the possibility of multi-dimensional (MD) behaviour raises several complexities both in terms of suspension phenomenology, material characterisation and solution of governing dynamics (Lester et al., 2010). First, the advent of MD dynamics means the stress state experienced by the particulate network is tensorial, involving a potential superposition of arbitrary shear and compressive stresses and strains (or strain rates). Despite a handful of limited studies (Islam and Lester, 2021; Lester et al., 2014), very little is known regarding the consolidation behaviour of particulate suspensions under such arbitrary loadings, let alone the development of general theory of consolidation or identification of controlling material properties. Second, although the hydrodynamics of the macroscopic suspension (specifically the phase-averaged suspension velocity field) are trivial in 1D (i.e. constant in space), in MD the suspension hydrodynamics must also be resolved via solution of the heterogeneous Navier-Stokes equations, which also requires characterisation of the suspension shear rheology.

Even under stagnant flow conditions, gravity separation in horizontal pipes is clearly a MD process due to the cylindrical shape of the pipe cross-section. This inherent multi-dimensionality can lead to the formation of a tensorial stress state (involving shear and compressive stress) in the settled bed that is currently not well understood. Similar to consolidation in vertical cylinders (Lester et al., 2014) there is potential for the shear yield stress in horizontal cylinders to retard consolidation via support of the network from the container walls, a problem which has not been explored for containers of any other shape. Furthermore, the cylindrical cross-section also raises the potential for the formation of gravity currents in the macroscopic suspension velocity field during sedimentation, leading to acceleration of sedimentation termed the Boycott effect (Boycott, 1920). Under the assumptions that such gravity currents and bulk suspensions flows are negligible and all solids concentration are aligned in the vertical direction, sedimentation in stagnant horizontal pipes may be described by an augmented 1D sedimentation theory (Landman and White, 1994) that accounts for the impact of the variable cross-sectional area of the pipe with horizontal depth. Note that these sedimentation assumptions are independent of those concerning the stress state of the settled bed, which can also deviate from 1D due to the cylindrical shape of the pipe and can retard consolidation of the bed by allowing the shear yield stress to support the compressive stress generated by the density difference between the phases.

A further assumption is that the material functions $P_y(\phi)$ and $R(\phi)$ are *constitutive* in that they correctly characterise suspension properties and so facilitate prediction of sedimentation and consolidation in other applications. While it is recognised (Buscall, 2009; Kim et al., 2007; Liétor-Santos et al., 2009; Manley et al., 2005; Lester et al 2013) that flocculated particulate suspensions are ratchet poro-elastic rather than plastic under compressive stress, the compressive yield stress $P_y(\phi)$ can be related to the bulk elastic modulus $K(\phi)$ as $P_y(\phi) = \int K(\phi)\, d\ln\phi$, and so characterisation by either material function is equivalent in light of irreversible consolidation. Similarly, in batch settling experiments, the hindered settling function $R(\phi)$ characterises the apparent sedimentation velocity of the suspension and thus the interphase drag. However, Rayleigh-Taylor instabilities are frequently observed during such tests due to local concentration fluctuations (Chandrasekhar, 2013; Drazin and Reid, 2004; Pan et al., 2001), leading to self-organisation of the sedimenting suspension into gravity current-driven vortex chains or formation of eroded channels that accelerate the sedimentation rate. As such, the sedimentation velocity and associated drag coefficient $R(\phi)$ estimated from such experiments are *apparent* quantities and thus it is unclear to what extent they can be used to predict sedimentation and consolidation in other scenarios. In this study, we shall determine to what extent such properties can be used to predict sedimentation and consolidation in horizontally oriented cylindrical pipes under stagnant flow.

**2.2 1D Batch settling theory**

The material functions $P_y(\phi)$ and $R(\phi)$ can be estimated from analysis of the solid-liquid interface profile obtained from gravity batch settling tests in vertical cylinders with a constant cross-sectional area. These tests are analysed as a 1D process where the solid particles settle while the balance liquid undergoes buoyant flow upwards with zero net suspension flux (Landman and White, 1994). The average solids phase velocity, $u(\phi)$, is assumed to be function of $\phi$ only and is related to the hindered settling function as

$$R(\phi) = \frac{\Delta\rho g\,(1-\phi)^2}{u(\phi)} \tag{1}$$

where $\Delta\rho$ is the difference between solid and liquid densities (kg/m$^3$) and $g$ is gravitational acceleration (m/s$^2$). From a gravity batch settling test the solids flux function, defined as $f(\phi) = \phi\, u(\phi)$, can be estimated as function of $\phi$ via the theory of Kynch (1952) which is based on the

conservation law for sedimenting suspensions at solids concentrations below the gel point ($\phi<\phi_g$)

$$\frac{\partial \phi}{\partial t} + \frac{\partial}{\partial x} f(\phi) = 0, \qquad (2)$$

$$\phi(x, 0) = \phi_0 \qquad 0 < x < h_0,$$

$$\phi(h_0, t) = 0 \qquad t > 0.$$

Here $x$ is the solid-liquid interface bed height in the vertical coordinate (m), $h_0$ is the initial suspension height (m) and $t$ is time (s). Under the assumption that the flux function, $f(\phi)$, has only one inflection point (which corresponds to common correlations such as Richardson and Zaki (1997) for the hindered settling function $R(\phi)$, $f(\phi)$ can then be estimated from evolution of the solid-liquid interface height (termed a settling curve) $h(t)$ over the range of solids concentrations expressed in the test up to the gel point $\phi_g$ (Diehl, 2007; Lester et al., 2005). At the gel point the suspension forms a particulate network that can transmit and resist compressive stress, leading to finite values for the compressive yield stress, i.e. $P_y(\phi) > 0$ for $\phi > \phi_g$, and $P_y(\phi) = 0$ for $\phi < \phi_g$. The associated solids diffusivity function $D(\phi)$ is then defined as

$$D(\phi) = \frac{(1-\phi)^2}{R(\phi)} \frac{dP_y}{d\phi}, \qquad (3)$$

which characterises the diffusion of solids stress throughout the network. Evolution of the solids concentration via sedimentation and consolidation during a batch settling experiment is then described by the 1D conservation equation subject to the initial and boundary conditions (Bürger et al., 2013; Buscall and White, 1987; Landman and White, 1994)

$$\frac{\partial \phi}{\partial t} + \frac{\partial}{\partial x} f(\phi) - \frac{\partial}{\partial x}\left[D(\phi)\frac{\partial \phi}{\partial x}\right] = 0, \qquad (4)$$

$$\phi(x, 0) = \phi_0 \qquad 0 < x < h_0,$$

$$\phi(h_0, t) = 0 \qquad t > 0,$$

$$\left.\frac{\partial \phi}{\partial x}\right|_{x=0} = \left.\frac{f(\phi)}{D(\phi)}\right|_{x=0} \qquad t > 0,$$

where the boundary condition at the bottom of the cylinder ($x = 0$) arises from the zero solids flux condition. Note that several studies (Diehl, 2007) have developed methods to also estimate the settling flux $f(\phi)$ at solids concentrations above the gel point. Eq. 4 is valid under the assumption that wall adhesion effects (Lester et al 2013, Lester and Buscall 2014) (where the suspension shear yield stress acts to support the suspension in the vertical cylinder) are negligible. These effects typically scale as $\tau_y(\phi)/R$ where $\tau_y(\phi)$ is the suspension shear yield stress and $R$ is the cylinder radius and so may be significant in narrow (radius <100mm) cylinders. Under steady-state conditions, Lester and Buscall (2014) develop a radially averaged 1D solids stress balance for the equilibrium solids profile $\phi(x,t) \to \phi_\infty(x)$ as $t \to \infty$

$$\Delta \rho g \phi_\infty(x) + P_y'(\phi_\infty(x))\frac{d\phi_\infty(x)}{dx} + \frac{2\tau_y(\phi_\infty(x))}{R} = 0. \tag{5}$$

Following the generalised transient solids concentration evolution equation (Lester, 2010) for colloidal suspensions

$$\frac{\partial \phi}{\partial t} + \langle \mathbf{u} \rangle \cdot \nabla \phi = \nabla \cdot \left( \frac{(1-\phi)^2}{R(\phi)} \left\{ \Delta \rho \, \phi \left( \frac{D\langle \mathbf{u} \rangle}{Dt} - \mathbf{g} \right) - \nabla \cdot \mathbf{\Sigma}_N \right\} \right), \tag{6}$$

where the network stress tensor $\mathbf{\Sigma}_N$ contains contributions from both the compressive and shear yield stresses, then the transient analogue of Eq. 4 is then

$$\frac{\partial \phi}{\partial t} + \frac{\partial}{\partial x}(f(\phi) + b(\phi)) - \frac{\partial}{\partial x}\left[D(\phi)\frac{\partial \phi}{\partial x}\right] = 0, \tag{7}$$

$\phi(x,0) = \phi_0 \qquad\qquad 0 < x < h_0,$

$\phi(h_0, t) = 0 \qquad\qquad t > 0,$

$\left.\frac{\partial \phi}{\partial x}\right|_{x=0} = \left.\frac{f(\phi)+b(\phi)}{D(\phi)}\right|_{x=0} \qquad t > 0,$

where $b(\phi) = \frac{(1-\phi)^2}{R(\phi)}\frac{2\tau_y(\phi)}{R}$. Following Lester and Buscall (2014), the shear $\tau_y(\phi)$ and compressive $P_y(\phi)$ yield stresses and the gel point $\phi_g$ can be estimated from the equilibrium solid-liquid interface height ($h_\infty$) data over a series of batch settling tests with different initial suspension heights ($h_0$) and/or initial solids volume fractions ($\phi_0$). The compressive yield stress is often modelled with the functional form (Landman et al., 1988)

$$P_y(\phi) = k\left(\left(\frac{\phi}{\phi_g}\right)^n - 1\right), \qquad (8)$$

where $k, n$ are fitting parameters. Based on this functional form, Buscall (2009) proposes the following functional form for the shear yield stress

$$\tau_y(\phi) = \frac{k\left(\left(\frac{\phi}{\phi_g}\right)^n - 1\right)}{\left(\frac{1}{S_\infty}-1\right)\left(1-\left(\frac{\phi}{\phi_g}\right)^{-n}\right)+1}, \qquad (9)$$

where $S_\infty$ is an additional fitting factor that characterises $\tau_y(\phi)$. Note that this functional form imposes the observed behaviour that $\tau_y(\phi)$ is of similar magnitude as $P_y(\phi)$ for solids concentrations near the gel point, but for $\phi > \phi_g$, the ratio $\tau_y(\phi)/P_y(\phi) \to S_\infty$. By approximating the steady force balance in Eq. 10 as

$$\Delta\rho g\, \Phi\left(p_{N,\infty}(x)\right) + \frac{dp_{N,\infty}(x)}{dx} + \frac{2S_\infty}{R}(p_{N,\infty}(x) + k) = 0, \qquad (10)$$

where $\Phi(P_y)$ is the inverse function of $P_y(\phi)$, we obtain the equilibrium pressure profile equation

$$x = \int_0^{p_{N,\infty}(x)} \frac{1}{\Delta\rho g\, \Phi\left(p_{N,\infty}(x)\right) + \frac{2S_\infty}{R}(p_{N,\infty}(x)+k)}\, dp, \qquad (11)$$

which upon integration yields the equilibrium pressure and solids concentration profiles

$$p_{N,\infty}(x) = k\left[(q + (1-q)e^{-rx})^{\frac{n}{n-1}} - 1\right], \quad \phi_\infty(x) = \phi_g(q + (1-q)e^{-rx})^{\frac{1}{n-1}}, \qquad (12)$$

where $q \equiv \frac{\Delta\rho g \phi_g R}{2S_\infty k}$, $r \equiv \frac{n-1}{n}\frac{2S_\infty}{R}$. Hence the equilibrium interface height ($h_\infty$) as a function of the "linear mass" is

$$m(h_\infty) = \phi_0\, h_0 = \int_0^{h_0} \phi_0\, dx = \int_0^{h_\infty} \phi_\infty(x)\, dx, \qquad (13)$$

which, upon insertion of Eq.12 yields the function

$$h_\infty(m) =$$

$$\frac{\phi_g}{r}\frac{n-1}{q-1}\left[-{}_2^1F_1\left(1,1,\frac{n-2}{n-1},\frac{q}{q-1}\right) + \left(1 + q(e^{rh_\infty}-1)\right)\left(q+(1-q)e^{rh_\infty}\right){}_2^1F_1\left(1,1,\frac{n-2}{n-1},\frac{qe^{rh_\infty}}{q-1}\right)\right], (14)$$

where ${}_2^1F_1(a,b,c,z)$ is the ${}_2^1F_1$ hypergeometric series. Hence fitting of Eq. 14 to equilibrium height data ($m$ vs $h_\infty$) facilitates estimation of the gel point $\phi_g$ and the shear and compressive yield stress parameters $k, n, S_\infty$. However, as Eq.14 involves four fitting parameters and equilibrium height data is relatively sparse, there exists a significant risk of spurious regression. To address this shortcoming, we also incorporate direct measurements of the suspension shear yield stress $\tau_y(\phi)$ to constrain the fitting of these parameters, as shall be discussed in Section 3.2.2 and 4.2.2. As such, $R(\phi)$ and $P_y(\phi)$ can be accurately estimated from a combination of shear yield stress tests and batch settling tests in vertical cylinders, even in the presence of significant wall adhesion effects. In this study we will only focus on characterisation of $R(\phi)$ under sedimentation only (i.e. for $\phi<\phi_g$).

## 2.3 1D Settling in arbitrary shaped containers

For arbitrarily shaped containers such as the horizontally oriented cylinder, wall effects in convex sections of the container may be more significant than that for vertical walls. However, the inclusion of wall adhesion into the solids stress balance (even at equilibrium) is beyond the scope of this study as it involves solution of the governing 2D hyperbolic stress equations (Lester et al, 2013) in the cylindrical geometry, currently an open problem. As discussed in Subsection 2.1, sedimentation and consolidation in containers of arbitrary shape is an inherently MD process that involves the potential for suspension gravity currents and tensorial network stress states. Under the assumption that both these effects (outlined in Section 2.1) and wall adhesion effects are negligible, several workers (Bürger et al., 2004; Landman et al., 1988; Usher and Scales, 2005) have extended the 1D model (Eq. 4) to account for the variable cross-sectional area $A(x)$ in settling vessels of arbitrary shape as

$$\frac{\partial}{\partial t}(A(x)\phi) + \frac{\partial}{\partial x}(A(x)f(\phi)) - \frac{\partial}{\partial x}\left[A(x)D(\phi)\frac{\partial \phi}{\partial x}\right] = 0 \text{ for } 0 < x < h_0, \qquad (15)$$

$\phi(x,0) = \phi_0 \qquad\qquad 0 < x < h_0,$

$\phi(h_0, t) = 0 \qquad\qquad t > 0,$

$\left.\frac{\partial \phi}{\partial x}\right|_{x=0} = \left.\frac{f(\phi)}{D(\phi)}\right|_{x=0} \qquad t > 0,$

For the case of a horizontal pipe of diameter $h_0$ and length $l$, the cross-sectional area $A(x)$ is given by the product of $l$ and the width $w(x)$ at height $0 < x < h_0$

$$A(x) = lw(x) = 2l\sqrt{\left(\frac{h_0}{2}\right)^2 - \left(x - \frac{h_0}{2}\right)^2} \qquad (16)$$

For sedimentation of suspensions (i.e. $\phi < \phi_g$), Eq. 15 simplifies to the conservation equation

$$\frac{\partial \phi}{\partial t} + \frac{\partial}{\partial x}f(\phi) = -\frac{A'(x)}{A(x)}f(\phi) \text{ for } 0 < x < h_0, \qquad (17)$$

hence the sedimentation is augmented by the shape of the container, as indicated by the source term in Eq. 17 above, where converging container walls (i.e. $A'(x) < 0$) act to accelerate sedimentation, and diverging container walls (i.e. $A'(x) > 0$) act to retard sedimentation. Application of steady-state conditions to Eq. 17 recovers the solids stress balance (Eq. 5), reflecting the assumption that shear stresses do not impact the solids concentration profile. Note however that for different container shapes, the relationship $h_\infty(m)$ in Eq. 14 is altered as the relationship $m \equiv \phi_0 h_0$ is no longer valid due to the changing cross-sectional area $A(x)$. Instead, the relationship between equilibrium height $h_\infty$ and the total volume V of solids in the experiment is given as

$$V(h_\infty) = \int_0^{h_0} \phi_0 A(x)\, dx = \int_0^{h_\infty} \phi_\infty(x) A(x)\, dx. \qquad (18)$$

In general, given estimation of the material functions $P_y(\phi)$ and $R(\phi)$, solution of Eq. 17 via numerical methods (outline in Section 3.2) can then be used to predict evolution of settling experiments in horizontally oriented pipes, including evolution of the settling curve $h(t)$ and the equilibrium height $h_\infty$.

## 3. Materials and methods

### 3.1 Experimental materials and methods

Several particulate suspensions were prepared which were comprised of Kaolin ASP 200 particles (from BASF supplied by Scott Chemicals Australia, Sauter mean diameter 2.68 μm and solids density 2650 kg/m$^3$) with various initial solid concentrations ($\phi_0$ = 1, 1.5 and 2.5 vol%) suspended in a 0.01 molar aqueous solution of KNO$_3$ to eliminate inconsistencies arising from variations in the ionic strength, and to ensure particle attraction by van der Waals forces by minimizing the electrostatic repulsion. Prior to sedimentation experiments, the suspensions were mixed at 100 rpm for 24 hrs ensure complete dispersion of particles.

*3.1.1 Experimental set up and procedure*

The vertical settling tests were conducted in 1L measuring cylinders with internal diameter 0.06 m and height 0.43 m. The kaolin suspensions were poured into the cylinder up to an initial height $h_0$ of around 0.092 cm and the solid-liquid interface height $h(t)$ over time $t$ was measured via a combination of visual observations and digital recording via a digital video camera. The equilibrium height $h_\infty$ was determined to be the solid-liquid interface height where there was no significant change over a 24-hour period. The horizontal settling tests were conducted in a horizontal cylindrical pipe segment of internal diameter 0.092 m and length 0.5 m. Kaolin suspensions were poured into the cylindrical pipe and the decrease in the solid-liquid interface height ($x$) was recorded over time ($t$) via a combination of visual observations and digital recording via a digital video camera. As shown in Figure 1, the horizontal pipe was placed inside a square water tank to minimise optical distortion from the curved cylinder surface and the entire horizontal experimental set up was placed inside a darkened environment to avoid optical reflections. Measurements confirmed that the solid-liquid interface height measured from outside the water tank scaled linearly with the interface height at the centre of the cylinder, and a linear correlation was developed to convert between these heights.

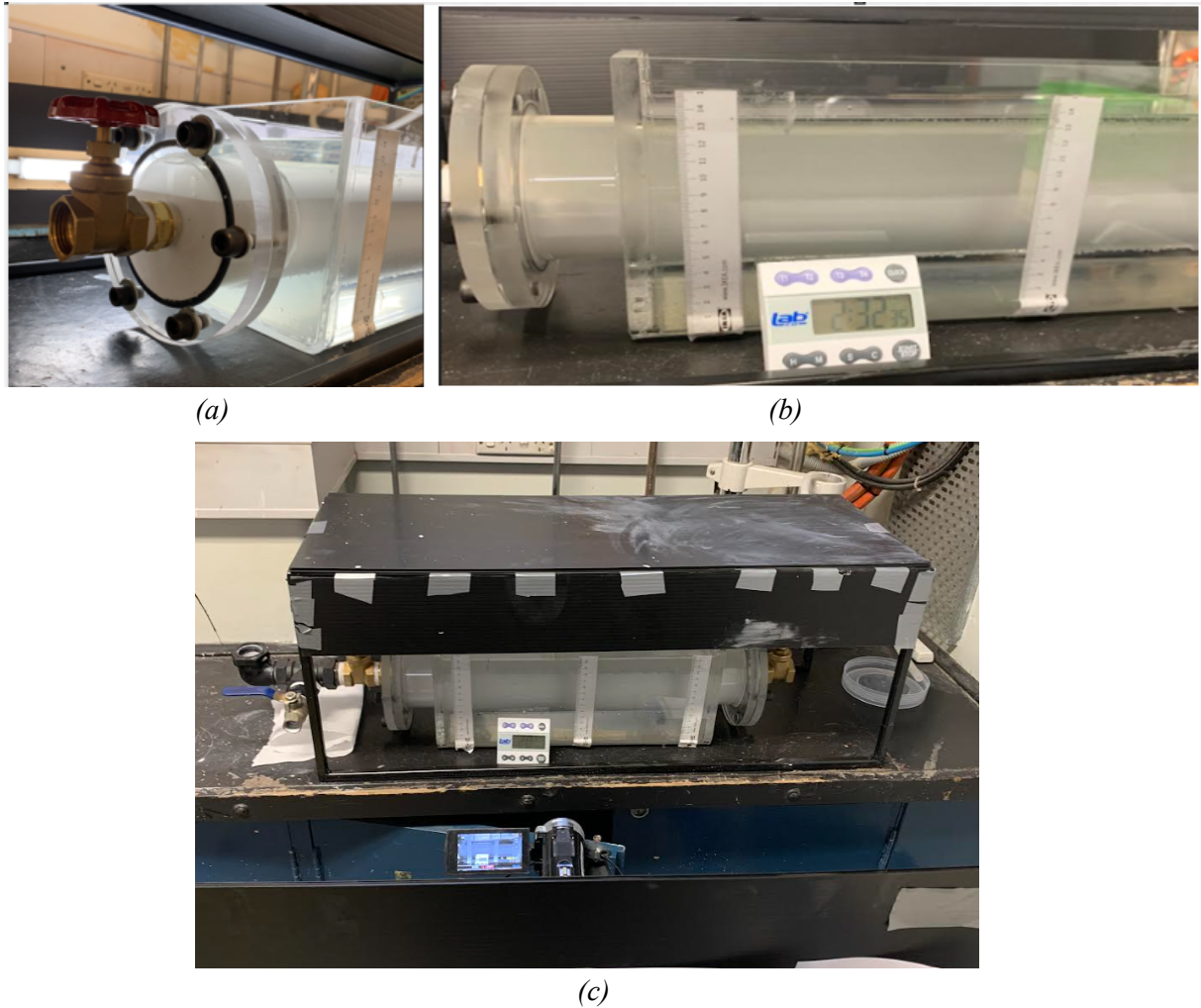

Figure 1: The set up for settling experiments in horizontal cylindrical pipe: (a) and (b) horizontal cylindrical pipe inside the water tank to prevent interference due to curvature of the pipe and (c) the entire set up with the shed to prevent reflection on the curved surface and video camera recording

### 3.2.2 Characterization of suspension properties

As outlined in Section 2.2, the hindered settling function $R(\phi)$ can be estimated from transient interface data, and the shear $\tau_y(\phi)$ and compressive $P_y(\phi)$ yield stresses can be estimated from a combination of shear yield stress measurements and equilibrium interface data. The hindered settling function $R(\phi)$ was estimated from the transient interface height $h(t)$ data measured during a vertical batch settling test at $\phi_0 = 1$ vol% and $h_0 = 0.35$ m in vertical cylinders and pressure filtration tests (20 – 600 kPa). The $R(\phi)$ values were extracted from the batch settling data using the methods outlined by Lester et al. (2005). To provide estimates of the hindered settling function at solids concentrations above the gel point, this data was combined with measurements of $R(\phi)$ for Kaolin suspensions via pressure filtration experiments (Stickland et al. (2008), which yields estimates of $R(\phi)$ in the range $\phi$=[0.2, 0.34]. The solids volume fraction gel point $\phi_g$ and the shear and compressive yield stress parameters $k, n, S_\infty$ were estimated by simultaneously fitting Eq. 14 and Eq. 9 to shear yield stress data from vane rheometer experiments (Stickland et al. (2008) and from the equilibrium height $h_\infty$ measurements for the vertical settling tests at different $\phi_0$ (1-2.5 vol%) and $h_0$ (0.01-0.2 m) as described in Section 2.2. The vane yield stress was measured using the vane-in-a-cup method (Dzuy and Boger, 1983). A four-bladed vane (15 mm diameter and 50 mm height) was submerged into the Kaolin ASP 200 suspension in a cup (45 mm diameter) and rotated at a controlled angular velocity of 0.2 rpm using a Haake VT550 viscometer (Thermo Fisher Scientific, Australia). The shear yield stress was calculated from the maximum torque (Hassan et al., 2025). Details of the extraction and curve fit for $R(\phi)$ and $P_y(\phi)$ as functions of solids volume fractions are outlined in Section 4.2.

### 3.2. Numerical Methods

*3.2.1 Numerical simulation of settling in vertical cylinders with wall effects*

For the simulation of settling in vertical cylinders with wall effects, a simple explicit finite difference method was used to solve the governing Eq. 17. A 1D mesh comprising of a single column of $J$=200 cells was used with a constant cell height of $\Delta x = h_0/J$, where $h_0$ is the initial suspension height. The time step ($\Delta t$) was chosen for the numerical simulation from the CFL stability criterion

$$\Delta t < \frac{\Delta x}{\max|f'(\phi)|+\max|b'(\phi)|+\frac{maxD(\phi)}{\Delta x}} \tag{19}$$

We reformulate equation (7) as

$$\frac{\partial \phi}{\partial t} + \frac{\partial}{\partial x}\left(f(\phi) + b(\phi) - \frac{\partial}{\partial x}d(\phi)\right) = 0 \text{ for } 0 < x < h_0, \tag{20}$$

$$\phi(x,0) = \phi_0 \qquad\qquad 0 < x < h_0,$$

$$\phi(h_0, t) = 0 \qquad\qquad t > 0,$$

$$\left.\frac{\partial \phi}{\partial x}\right|_{x=0} = \left.\frac{f(\phi)+b(\phi)}{D(\phi)}\right|_{x=0} \qquad t > 0,$$

where $d(\phi) \equiv \int_0^\phi D(\phi)\,d\phi$. Equation (20) is then discretized as

$$\varphi_{n+1,j} = \varphi_{n,j} - \frac{\Delta t}{\Delta x}\left(F_{n,j+\frac{1}{2}} - F_{n,j-\frac{1}{2}} + B_{n,j+\frac{1}{2}} - B_{n,j-\frac{1}{2}} + D_{n,j+\frac{1}{2}} - D_{n,j-\frac{1}{2}}\right) = 0, \tag{21}$$

with no-flux boundary conditions

$$\varphi_{n+1,1} = \varphi_{n,1} - \frac{\Delta t}{\Delta x}\left(F_{n,\frac{3}{2}} + B_{n,\frac{3}{2}} + D_{n,\frac{3}{2}}\right) = 0, \tag{22}$$

$$\varphi_{n+1,J} = \varphi_{n,J} - \frac{\Delta t}{\Delta x}\left(-F_{n,J-\frac{1}{2}} - B_{n,J-\frac{1}{2}} - D_{n,J-\frac{1}{2}}\right) = 0.$$

where $\varphi_{n,j} \equiv \varphi(x = (j - 1/2)\,\Delta x, t = n\,\Delta t)$, $\phi_{n,j} \equiv \phi(x = (j - 1/2)\,\Delta x, t = n\,\Delta t)$, with j=1:J and $F$, $B$, $D$ respectively are the convective and diffusive fluxes

$$F_{n,j+\frac{1}{2}} = f_1(\phi_{n,j}) + f_2(\phi_{n,j+1}), \tag{23}$$

$$B_{n,j+\frac{1}{2}} = b(\phi_{n,j}),$$

$$D_{n,j+\frac{1}{2}} = -\frac{d(\phi_{n,j+1}) - d(\phi_{n,j+1})}{\Delta x},$$

where the convective fluxes in (6) are handled via an Enquist-Osher flux splitting scheme with

$$f_1(\phi) = \begin{cases} 0 & \text{if } \phi < \phi_i \\ f(\phi) - f(\phi_i) & \text{if } \phi > \phi_i \end{cases}, \qquad f_2(\phi) = \begin{cases} f(\phi) & \text{if } \phi < \phi_i \\ f(\phi_i) & \text{if } \phi > \phi_i \end{cases}, \tag{24}$$

where $\phi_i$ is the inflection point of the flux function $f(\phi)$. The total solids volume is conserved as

$$\int_0^{h_0} \phi_0\,dx = \int_0^{h_\infty} \phi_\infty(x)\,dx \approx \Delta x \sum_{j=1}^{J} \phi_{n,J}. \tag{25}$$

*3.2.2 Numerical simulation of settling in horizontal pipes without wall effects*

For the simulation of settling in horizontal cylindrical pipes, a simple explicit finite difference method similar to that used for the vertical settling simulations was used to solve the governing Eq. 17. Again, a 1D mesh comprising of a single column of *J*=200 cells was used with a constant cell height of $\Delta x = h_0/J$, where the initial suspension height $h_0$ is equal to the pipe diameter. The time step ($\Delta t$) was chosen for the numerical simulation from the CFL stability criterion

$$\Delta t < \frac{\Delta x}{\max|f'(\phi)| + \frac{maxD(\phi)}{\Delta x}} \tag{26}$$

As equation (6) is conservative with respect to the product $\varphi(x,t) \equiv A(x)\phi(x,t)$, we reformulate this equation as

$$\frac{\partial \varphi}{\partial t} + \frac{\partial}{\partial x}\left(A(x)f(\phi) - A(x)\frac{\partial}{\partial x}d(\phi)\right) = 0 \text{ for } 0 < x < h_0, \tag{27}$$

$$\phi(x,0) = \phi_0 \qquad 0 < x < h_0,$$

$$\phi(h_0, t) = 0 \qquad t > 0,$$

$$\left.\frac{\partial \phi}{\partial x}\right|_{x=0} = \left.\frac{f(\phi)}{D(\phi)}\right|_{x=0} \qquad t > 0,$$

Where $d(\phi) \equiv \int_0^\phi D(\phi)\, d\phi$. Equation (16) is then discretized as

$$\varphi_{n+1,j} = \varphi_{n,j} - \frac{\Delta t}{\Delta x}\left(F_{n,j+\frac{1}{2}} - F_{n,j-\frac{1}{2}} + D_{n,j+\frac{1}{2}} - D_{n,j-\frac{1}{2}}\right) = 0, \tag{28}$$

With boundary conditions

$$\varphi_{n+1,1} = \varphi_{n,1} - \frac{\Delta t}{\Delta x}\left(F_{n,\frac{3}{2}} + D_{n,\frac{3}{2}}\right) = 0, \tag{29}$$

$$\varphi_{n+1,J} = \varphi_{n,J} - \frac{\Delta t}{\Delta x}\left(-F_{n,J-\frac{1}{2}} - D_{n,J-\frac{1}{2}}\right) = 0.$$

where $\varphi_{n,j} \equiv \varphi(x = (j-1/2)\Delta x, t = n\Delta t)$, $\phi_{n,j} \equiv \phi(x = (j-1/2)\Delta x, t = n\Delta t)$, with j=1:J and $F, D$ respectively are the convective and diffusive fluxes

$$F_{n,j+\frac{1}{2}} = A\left[\left(j + \tfrac{1}{2}\right)\Delta x\right]\left[f_1(\phi_{n,j}) + f_2(\phi_{n,j+1})\right], \tag{30}$$

$$D_{n,j+\frac{1}{2}} = -\frac{1}{\Delta x}A\left[\left(j + \tfrac{1}{2}\right)\Delta x\right]\left[d(\phi_{n,j+1}) - d(\phi_{n,j+1})\right].$$

The total solids volume is conserved as

$$V(h_\infty) = \int_0^{h_0} \phi_0\, A(x)\, dx = \int_0^{h_\infty} \phi_\infty(x) A(x)\, dx \approx \Delta x \sum_{j=1}^{J} \varphi_{n,J}. \tag{31}$$

## 4. Results and discussion

### 4.1 Vertical and horizontal batch settling tests

To directly compare results, the vertical and horizontal settling tests were conducted at an initial height of $h_0 = 0.092$m which is the equal to the diameter of the horizontal cylindrical pipe of 0.092 m. The solid-liquid interface profiles for the vertical and horizontal settling tests for the 1, 1.5 and 2.5 vol% suspensions are shown in Figure 2.

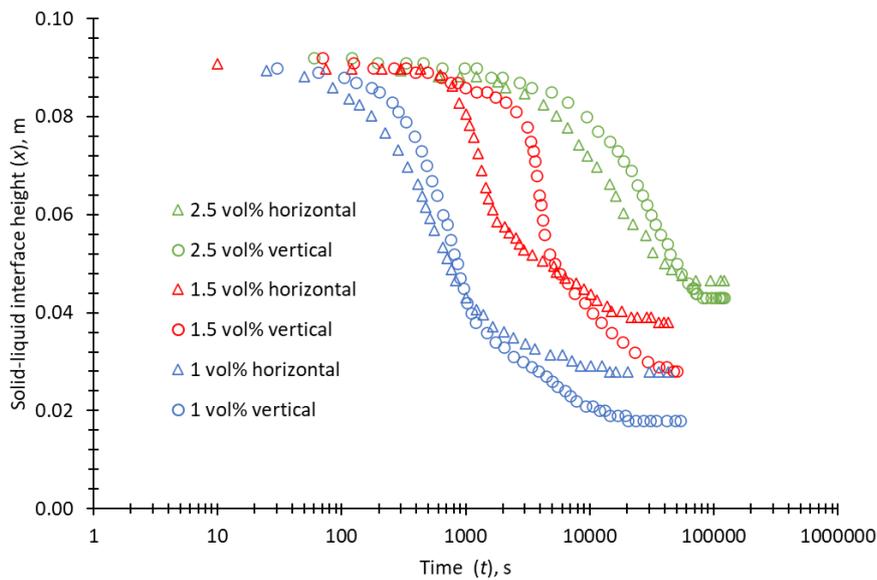

*Figure 2: Experimental solid-liquid interface vs time profile for the settling tests in the vertical cylinders and horizontal cylindrical pipe and experimental solid-liquid interface vs time profile for the settling tests in the vertical cylinder at $\phi_0$ = 1, 1.5 and 2.5 vol%*

At the beginning of the settling tests, channel formation and decay of internal flows retard the initial sedimentation process, leading to characteristic S-shaped settling curves (Lester et al., 2005). We invoke the assumption that these initialization mechanisms do not influence the settling behaviour except to delay attainment of a constant settling velocity regime. Figure 2 indicates that at the same solids volume fraction, the vertical cylinder tests appear to have longer initialization effects the horizontal cylinder tests. For all cases, the initial sedimentation rate is faster in the horizontal cylindrical pipe than in the vertical cylinder, consistent with Eq. 17 and observations from previous studies (Anestis, 1981; Bürger et al., 2004). Conversely, as expected from Eq. 14, the equilibrium solid-liquid interface height for the horizontal cylinder tests is consistently higher than those for the vertical cylinder tests.

## 4.2 Characterisation of solid-liquid separation properties

*4.2.1 Characterization of the hindered settling function $R(\phi)$*

The hindered settling function $R(\phi)$ as a function of solids volume fraction $\phi$ was estimated from analysis of the vertical settling test with initial solids concentration $\phi_0 = 1$ vol% and $h_0 = 0.35$ m, which is combined with estimates of $R(\phi)$ from pressure filtration experiments on Kaolin suspensions (Stickland et al., 2008). Figure 3 shows the experimental and curve fit settling profile $h(t)$ for the vertical batch settling test with initial solids volume fraction $\phi_0 = 1$ vol% and initial height $h_0 = 0.35$ m.

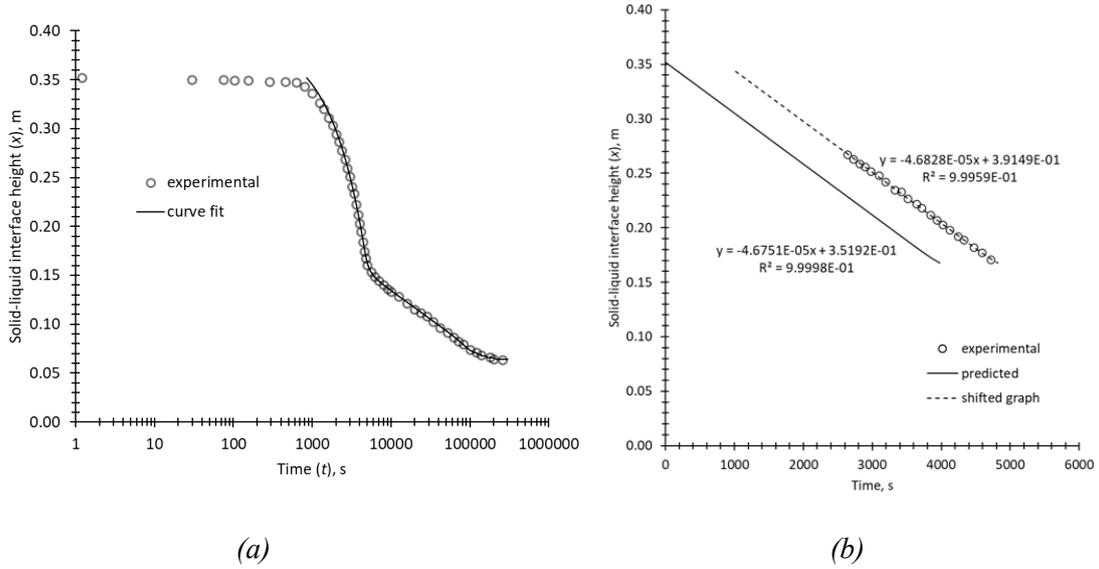

*Figure 3: (a) Solid-liquid interface height vs time profile and (b) shifted curve fit and experimental solid-liquid interface profile for the settling test in vertical cylinders for Kaolin ASP 200 suspension at $\phi_0=1$ vol% and $h_0=0.35$ m*

Following the approach described in (Lester et al., 2005), a time shift of approximately 842 s was applied to the predicted $h(t)$ curve as presented in Figure 3(a), after which linear settling behaviour is observed up to approximately $t = 3000$ s. The time shift in Fig. 3(a) was optimized so that the slope i.e. settling velocities of the experimental and fitted curves in the linear region (2000-3000 s) is identical (within 0.5%). The unshifted and shifted predicted solid-liquid interface profile and the experimental solid-liquid interface height vs time data used for this optimization are presented in Fig. 3(b). The initial settling rate, $u(\phi_0)$, is a function of the initial solids volume fraction $\phi_0$ (Coe and Clevenger, 1916; Kynch, 1952), and $u(\phi_0) = -h'(t) = 4.4024 \times 10^{-5}$ m/s corresponding to $R(\phi_0) = 3.6 \times 10^8$ kg/m³s. The $R(\phi)$ data in the range of $\phi_{inf} < \phi < \phi_{max}$ was extracted using the analytical method described in Lester et al. (2005). At $\phi_{inf} = 0.012$, $f'(\phi)$ is maximum or $f''(\phi) = 0$. At $\phi_{max} = 0.0347$ is the maximum analytic estimate of $R(\phi)$. To generate a continuous function for $0 < \phi < \phi_0$, $R(\phi)$ data was extrapolated so that both $R(\phi)$ and $R'(\phi)$ are continuous at $\phi = \phi_0$. However, it is to be noted that the entire suspension in the cylinder is at $\phi = \phi_0$ at the beginning of the batch sedimentation test and hence the $R(\phi)$ values below $\phi = \phi_0$ does not play any role in the numerical simulation. To develop a continuous function for solids volume fractions beyond gel point $\phi_g$, the $R(\phi)$ values extracted from the pressure filtration (Stickland et al., 2008) were used. The $R(\phi)$ values in the entire range of solids volume fractions including sedimentation and filtration region were curve fit as a fourth order interpolation function as presented in Fig. 4.

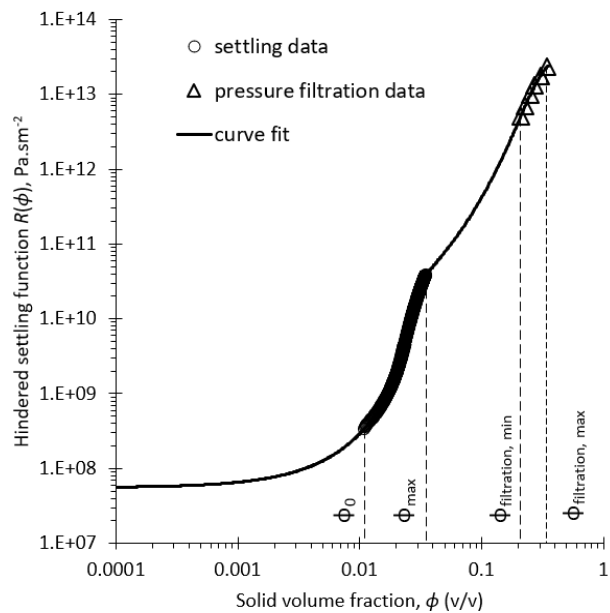

*Figure 4: Hindered settling function, $R(\phi)$ as continuous functions of solids volume fraction for Kaolin ASP 200 suspension extracted from the analysis of settling tests at $\phi_0=1$ vol% and $h_0 = 0.35$ m*

Fig. S.1(a)-(b) in supplementary section show the settling velocity $u(\phi)$ and solids flux $f(\phi)$ functions that correspond to the $R(\phi)$ function estimated from the pressure filtration data and the batch settling test at $\phi_0=1$ vol% and initial height $h_0 = 0.35$ m.

### 4.2.2 Characterization of the shear and compressive yield stress

Along with the shear yield stress data $\tau_y(\phi)$ at different solids concentrations $\phi$, the equilibrium solid-liquid interface heights $h_\infty$ from the vertical settling tests at different initial heights $h_0$ and initial solids volume fractions $\phi_0$ was used to estimate the suspension gel point $\phi_g$ and the parameters $k$, $n$, $S_\infty$. As shown in Figure 7(a), the linear mass $m = \phi_0 h_0$ was fitted as a function of the equilibrium height $h_\infty$ via equation (4), and in Figure 7(b) the shear yield stress $\tau_y(\phi)$ was fitted as a function $\phi$ via Eq. 9. This fitting was performed via a simultaneous minimisation of the sum of squared residuals, resulting in a coefficient of determination of $R^2=0.998$. This fitting yielded the parameter values $\phi_g$=4.196 vol%, $n$=4 [-], $k$=5.265 [Pa], $S_\infty$=0.080 [-]. The resultant shear and compressive yield stress functions are shown in Figure 5(a) and (b), the evolution of the ratio of the shear to compressive yield stress from unity to $S_\infty$ is shown in Figure 6 (a) and (b).

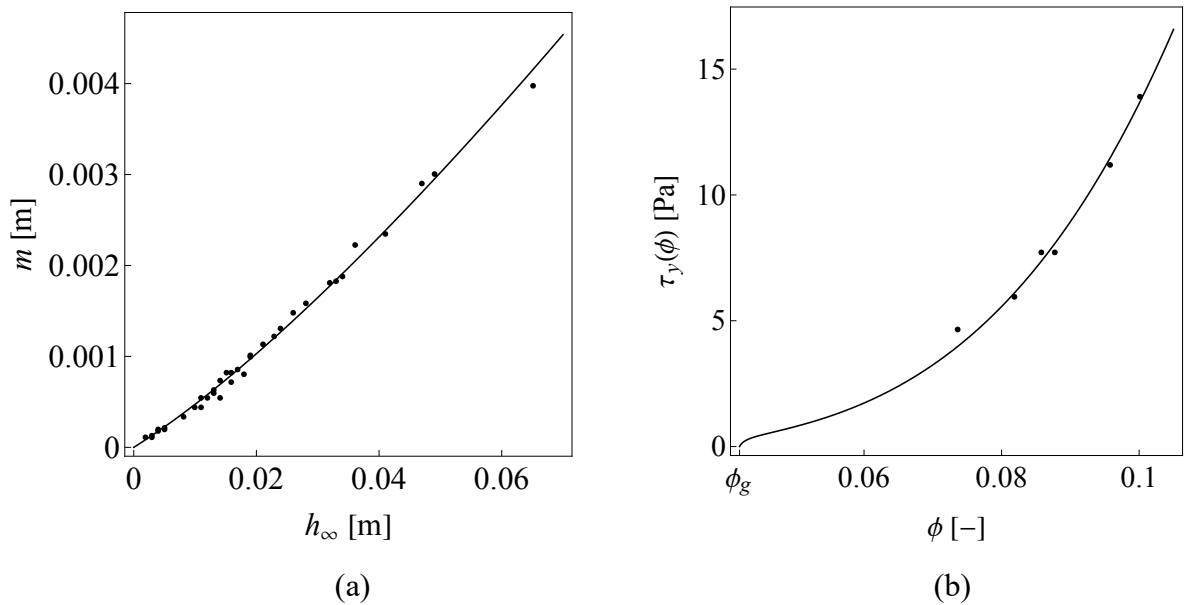

(a)          (b)

*Figure 5: (a) Plot of experimental data (points) and fitted (Eq. 14) (line) of linear mass m against equilibrium height in vertical settling tests. (b) Plot of experimentally measured suspension shear yield stress $\tau_y(\phi)$ against solids volume fraction $\phi$ data (points) and fitted (Eq. 9)(line).*

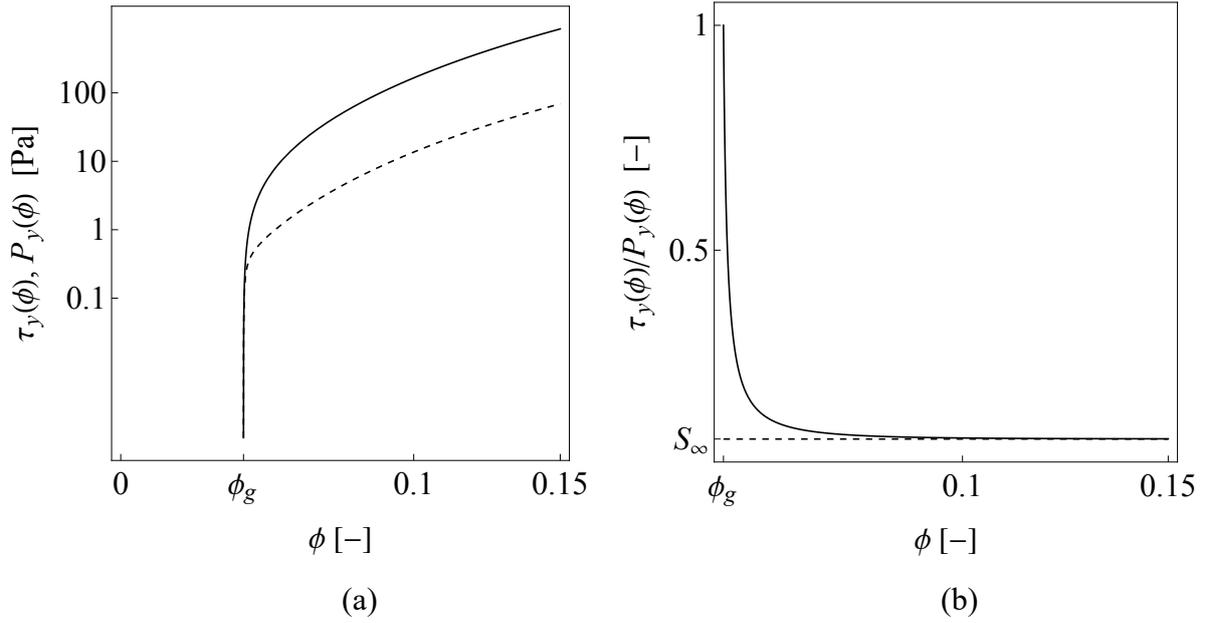

*Figure 6: (a) Fitted shear and compressive yield stress functions from fitting of data shown in Figure 5 and (b) Evolution of the ratio of shear to compressive yield stress from unity at the gel point to S at higher solids concentration.*

### 4.2.3 Verification of suspension sedimentation and consolidation

To verify that the hindered settling function and compressive yield stress were accurately estimated from the vertical batch settling data, the solid-liquid interface profile for the vertical settling test at $\phi_0 = 1$ vol% and $h_0 = 0.35$ m was curve fit from the 1D Eq. 7 in Section 2.2. A comparison of the experimental and numerical solid-liquid interface profiles was presented in Fig., showing excellent agreement (as expected) with errors of less than 3% (Fig. S.2(b) in supplementary section) in $h(t)$ for all times once the initialization dynamics have decayed and the linear settling regime is established. Beyond the linear settling region, the solid-liquid interface height vs time profile becomes nonlinear up to $t = 100000$ s corresponding to $\phi_g = 4.2$ vol% beyond which compression becomes dominant. The solids volume fractions at the solid-liquid interface over time are presented in Fig. S.2(a) in supplementary section showing the onset of compression at around $t = 100000$s (corresponding to $\phi_g = 4.2$ vol%).

The percentage error between the experimental and predicted (shifted to exclude the initialization region) solid-liquid interface height vs time data is presented in Fig. S.2(b) in supplementary section. In general, the percentage error is within ± 3%. Beyond the initialization region and before the onset of compression, the settling is dominated by the hindered settling function. Comparison of the experimental and curve fit solid-liquid interface data in Fig. 3(a) and Fig. S.2(a) beyond the initialization region ($t\sim850$ s) up to onset of compression effects ($t\sim100000$ s), provided accuracy of the estimation of $R(\phi)$ in the solids volume fraction range of relevant to hindered settling only ($\phi_0 = 1$ vol% to $\phi_g = 4.2$ vol%).

When the solids volume fraction at the interface reaches the gel point $\phi_g = 4.2$ vol%, the compressive yield stress, $P_y(\phi) > 0$, and the settling profile $h(t)$ is controlled by the combined effect of hindered settling and compression (consolidation), as is described by Eq. (4). As the gel point, $\phi_g$ and the compressive yield stress, $P_y(\phi)$ were estimated from the equilibrium height data from vertical settling tests, accurate estimation of the solid-liquid interface profile beyond $\phi_g$ verifies the technique applied to estimate $\phi_g$ and $P_y(\phi)$.

### 4.3 Prediction of settling in vertical cylinders

A comparison of the experimental and predicted interface height $h(t)$ profiles for the vertical settling tests over $\phi_0 = 1$, 1.5 and 2.5 vol% are presented in Fig. S.3 in the supplementary section. The predicted $h(t)$ curves were generated from numerical solution of Eq. 3 and Eq. 4. The sedimentation initialization effects occur up to t ~ 200 s, 3000s and 7000s respectively for $\phi_0 = 1$, 1.5 and 2.5 vol% tests, after which linear settling region occurs. As shown in Fig. S.5(a), the time shift applied to the experimental data to remove initialisation effects is a linearly increasing function of solids concentration for the vertical settling tests. For all tests the predicted settling profile matched the experimental one very well (within ± 5%), except for the $\phi_0 = 1$% test in the range of 3000-18000s, indicating accuracy of the estimated $R(\phi)$ values. For the $\phi_0$ = 1, 1.5 and 2.5 vol% tests, the solids volume fraction at the interface reaches the gel point, $\phi_g$ at approximately 18000, 28000 and 50000s respectively, indicating the onset of compression effects as shown in Fig. S.4(a)-(c). The time after which the compression effect had an impact on the interface profile is presented in Fig. S.5(b) as a linear function of solids volume fraction ($\phi$). As presented in Fig. S.4(d), the precited and experimental equilibrium heights are within ±5%; however, the analysis of the pressure gradient, compressive yield stress and shear stress components for the solids volume fraction profile at equilibrium height indicated that the shear stress component (wall effects) is approximately ±10% of combined effect of pressure gradient and compressive yield stress. Despite these moderate effects of shear stress components or wall effect in vertical cylinder tests, these results indicate that the settling properties estimated from the $\phi_0$ = 1 vol% vertical test can be used accurately predict settling in the $\phi_0$ = 1, 1.5 and 2.5 vol% vertical tests.

A summary of experimental and predicted equilibrium solid-liquid interface height ($h_\infty$) for the settling tests in vertical cylinders is provided in Table 1. The percentage error for the predicted equilibrium interface height is within ±5% of the experimental value.

*Table 1: Experimental and predicted equilibrium solid-liquid interface height ($h_\infty$) for the settling tests in vertical cylinders at $\phi_0$ = 1, 1.5 and 2.5 vol%*

| Solids volume fraction ($\phi$), vol% | Equilibrium solid-liquid interface height ($h_\infty$), m | | Error (%) |
|---|---|---|---|
| | Experimental | Predicted | |
| 1 | 0.01800 | 0.01855 | -3.05 |
| 1.5 | 0.02800 | 0.02677 | 4.40 |
| 2.5 | 0.04300 | 0.04080 | 5.12 |

### 4.4 Prediction of settling in horizontal cylindrical pipes

In this subsection we compare the measured and predicted settling profiles in the horizontally oriented cylinders. The experimental and predicted settling profiles $h(t)$ for the horizontal settling tests at $\phi_0$ = 1, 1.5 and 2.5 vol% are presented in Fig. 7. The predicted $h(t)$ profile for these settling tests were generated from the numerical simulation of the governing Eq. 17, and the estimated material functions $R(\phi)$ and $P_y(\phi)$ from the settling test in vertical cylinder at $\phi_0$ = 1 vol% and $h_0$ = 0.35 m as outlined in Section 4.2.

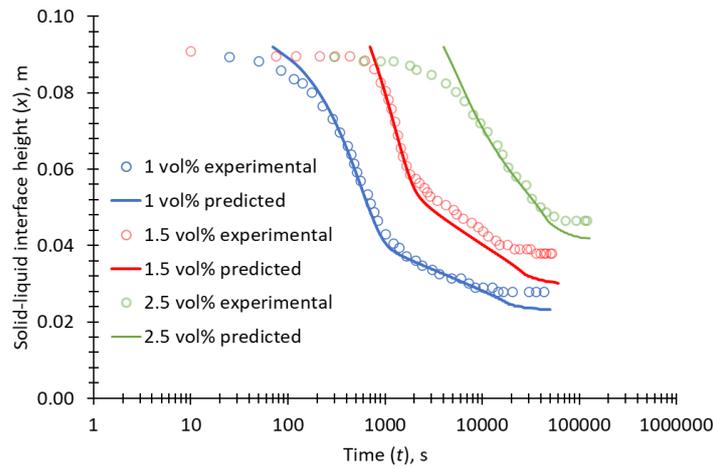

*Figure 7: Experimental and predicted solid-liquid interface vs time profile for the settling tests in the horizontal cylindrical pipe at $\phi_0$ = 1, 1.5 and 2.5 vol%*

*4.4.1 Initialization regime*

Similar to the vertical settling tests, time shifts of 120, 700 and 1500s were applied to the predicted interface profile for the $\phi_0$ = 1, 1.5 and 2.5 vol% tests respectively to match the experimental data in the sedimentation regime. As shown in Fig. 8(a), the time shift applied to the experimental data to remove initialisation effects is a second order polynomial function of solids concentration for the vertical settling tests. Combining Fig. S.5(a) and 8(a), the time shift applied for the horizontal settling tests can be predicted as a second order function of the time shift applied for the vertical settling tests as presented in Fig. 8(b).

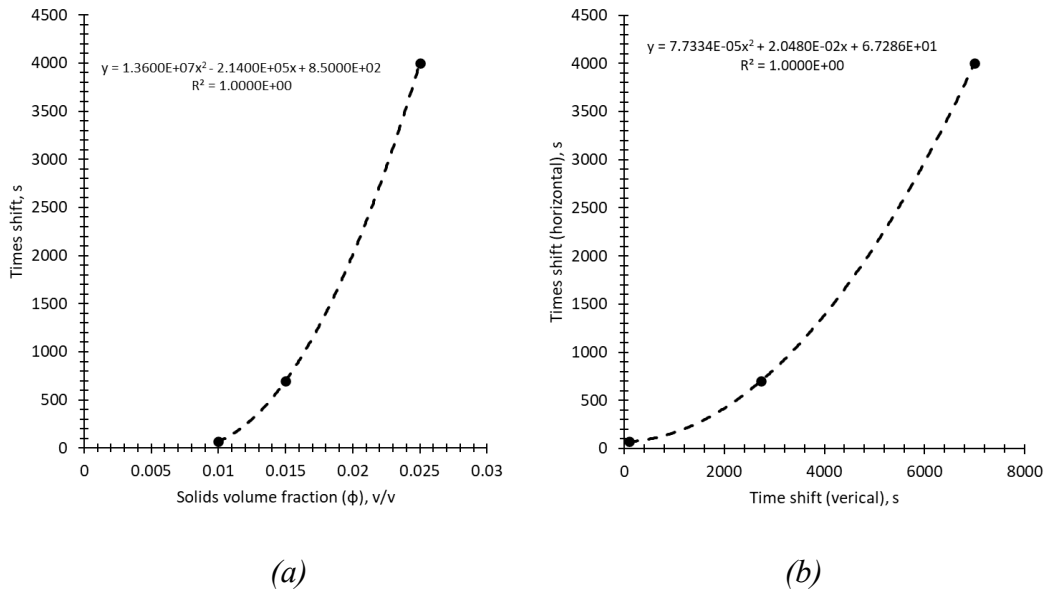

*Figure 8: (a)Time shift applied to the predicted interface profile for the settling tests in horizontal cylindrical pipes and (b) time shift applied to the predicted interface profile for the horizontal settling tests as a function of time shift applied to the predicted interface profile for the vertical settling tests at $\phi_0 = 1$, 1.5 and 2.5 vol% as second order function of solids volume fraction ($\phi$)*

*4.4.2 Sedimentation regime*

As can be observed from Fig.7, the predicted interface profile for the settling tests in horizontal cylindrical pipes matched well with the experimental interface profile for the three solids concentrations. This is also reflected in the percentage error vs time plot in Fig. S.6(d) in the supplementary section where the percentage error is within ±10% for the 1 and 1.5 vol% and within ±5% for the 2.5 vol% settling tests. This verifies the accuracy of the estimated $R(\phi)$ values estimated in Section 4.2 from the settling tests in vertical cylinder at $\phi_0 = 1$ vol% and the applicability of these values in predicting the interface profile in the sedimentation regime for the settling tests in horizontal cylindrical pipes. Overall, this indicates that the settling behaviour of the solid-liquid suspensions in horizontal cylindrical pipes in the sedimentation regime can be predicted from the material properties extracted from the settling tests conducted in vertical cylinders.

*4.4.3 Compression regime*

It is observed in Fig. 7 that the solid-liquid interface profile predicted from the numerical simulation generally underpredicts the experimental profile for the tests at three solids volume fractions in the compression region. From Fig. S.(a)-(c), the solids volume fraction reaches the gel point, $\phi_g$ at approximately 18000, 28000 and 50000 s indicating the onset of compression effects. A summary of experimental and predicted equilibrium solid-liquid interface height ($h_\infty$) is provided in Table 2, indicating significant errors (up to ~20%) between the predicted and measured equilibrium interface height.

*Table 2: Experimental and predicted equilibrium solid-liquid interface height ($h_\infty$) for the settling tests in horizontal cylindrical pipe at $\phi_0$ = 1, 1.5 and 2.5 vol%*

| Solids volume fraction ($\phi$), vol% | Equilibrium solid-liquid interface height ($h_\infty$), m | | Error (%) |
|---|---|---|---|
| | Experimental | Predicted | |
| 1 | 0.02792 | 0.02323 | 16.80 |
| 1.5 | 0.03795 | 0.03047 | 19.7 |
| 2.5 | 0.04648 | 0.04192 | 9.81 |

The significant nature of these errors suggest that the equilibrium stress state in the horizontal cylinder does not solely consist of gravitational stresses and compressive stresses, but wall effects are also significant. These errors are consistent with the finding in Section XX that wall effects account for ~11% of the compressive stress in vertical cylinder tests. Wall effects are expected to be more significant in the lower half of the horizontal cylinder than for the vertical cylinder, as shear stresses are maximal when the container wall is at 45 degrees to horizontal. This also explains why the error in Table 2 is maximal for $\phi_0$=1.5 vol%, as the corresponding equilibrium height is closest the 45 degree part of wall. Hence wall adhesion effects are significant in horizontally oriented cylinders. Unfortunately, solution of the stress state in the horizontally oriented cylinder is significantly more complex than that for the vertically oriented cylinder, and is beyond the scope of this study.

*4.4.4 Discussion*

From the results discussed in Section 4.4.2, it is evident that under stagnant conditions the separation behaviour of particulate suspensions in horizontal cylindrical pipes in the sedimentation regime can be accurately predicted from the sedimentation properties extracted from the analysis of settling experiments in vertical cylinder. This reflects the sedimentation properties extracted from the tests in vertical cylinders are accurate and representative material properties and the transient effects such as gravity currents are not significant in the settling of particulate suspensions in the horizontal cylindrical pipes. On the other hand, as discussed in Section 4.4.3, the separation behaviour in the horizontal cylindrical pipes in the consolidation regime is not accurately predicted by the 1D theory due to the significant wall adhesion effects. Although resolution of the shear and compressive stress state in horizontally oriented cylinders is beyond the scope of this study, future resolution of this stress state is expected to reconcile these differences.

The techniques developed in this study to characterize the settling behaviour of particulate suspension in the horizontal cylindrical pipes are more complex in nature compared to the analysis of vertical cylinder tests. These techniques have not been applied to estimate the settling properties yet, despite its direct relevance to industrial applications. The results from this study imply that the application of 1D sedimentation worked well in the sedimentation region but could not capture the consolidation behaviour for the horizontal pipes. Despite these shortcomings, these stagnant cylindrical pipe results provide a basis for the development of methods for predicting pipeline sedimentation under laminar and turbulent flow conditions.

## 5. Conclusion

In this study, batch settling tests in a horizontal cylindrical pipe with Kaolin ASP200 suspensions at initial solids volume fractions $\phi_0$ = 1, 1.5 and 2.5 vol% and initial height $h_0$ = 0.092 m were conducted to investigate the solid-liquid separation behaviour of suspensions in horizontal cylindrical pipes with varying cross-sectional area. The solid-liquid interface height vs time profiles for these tests were predicted from the numerical simulation using hindered settling function, $R(\phi)$ extracted from the analysis of batch settling tests at $\phi_0$ = 1 vol% and $h_0$ = 0.35 m in the vertical cylinder with constant cross-sectional area. The predicted solid-liquid interface height vs time profiles were compared to the experimental height vs time profile. Apart from the initialization and the compression regions, the experimental and predicted profile matched well in the linear and hindered settling regions of the settling curve. However, the profile obtained from numerical simulation in general underpredicted the experimental profile, with the difference being more significant in the compression region. For the three tests, the percentage error for the predicted equilibrium interface height, which is dominated by compression effects, was within 20% of the experimental value.

For the numerical simulation, the settling of solid-liquid suspensions in horizontal cylindrical pipes with varying cross-sectional area was assumed to be a 1D process with zero net flux. Comparison of the solid-liquid interface profiles from experimental observation and numerical simulation indicated that the estimation of solid-liquid behaviour in horizonal pipes is feasible without considering the flow in directions other than the vertical coordinate. In addition, the hindered settling function extracted from the analysis of settling tests in vertical cylinders were used for the prediction of the settling behaviour in horizontal pipes which is a complex system with limitations such as requirement of more sample and space and surface curvature causing difficulty in tracking the solid-liquid interface. Hence, numerical simulation of the conservation equations assuming the settling as an one dimensional process and utilising the material properties extracted from the analysis of settling in vertical cylinders is a convenient technique for estimating the solid-liquid separation behaviour of suspensions in horizontal cylindrical pipes.

## Acknowledgements

The authors acknowledge the financial support from Australian Research Council (ARC-LP180100869) and the logistic support from the ARC Centre of Excellence for Enabling Eco-Efficient Beneficiation of Minerals, The University of Melbourne for the experiments. The authors thank Yuxuan Luo, Raul Cavalida and Sajid Hassan for their assistance in carrying out the experiments.